\documentclass[useAMS,usenatbib,usegraphicx]{mn2e}

\title[Nuclear activity in ellipticals]{Nuclear activity and the dynamics of elliptical galaxies}
\author[M.R. Merrifield]
{M.R.\ Merrifield$^1$\thanks{E-mail: michael.merrifield@nottingham.ac.uk}\\
$^1$School of Physics \& Astronomy, University of Nottingham, 
    University Park, Nottingham, NG7 2RD
}
\begin{document}

\date{Accepted 2004 ????? ??. Received 2004 ?????? ??; in original form 2004 March 30}

\pagerange{\pageref{firstpage}--\pageref{lastpage}} \pubyear{2002}

\maketitle

\label{firstpage}

\begin{abstract}
This paper looks for any correlation between the internal dynamics of
elliptical galaxies and the relatively mild nuclear activity found in
many such systems.  We show that there is such a relation in the sense
that the active ellipticals tend to be significantly less rotationally
supported than their inactive cousins.  The correlation can partly be
related to the galaxies' luminosities: the brightest galaxies tend to
be more active and less rotationally supported.  However, even at
lower luminosities the active and inactive galaxies seem to have
systematically different dynamics.  This variation suggests that there
are significant large-scale structural differences between active and
inactive elliptical galaxies, and hence that the existence of both
types of system cannot just be the result of random sporadic nuclear
activity.
\end{abstract}


\section{Introduction}\label{sec:intro}
 Perhaps the most basic remaining question in the study of active
galactic nuclei (AGN) is why some nuclei are active while others are
not.  Now that it is well established that massive black holes lie at
the centres of the vast majority of galaxies \citep{kr95}, we know
that it cannot be the lack of the necessary power plant, which only
really leaves the fuel supply.  The simplest explanation then comes
down to the duty cycle of the fueling process: in this scenario, all
galaxies are active from time to time, and the ones that are observed
as AGN are simply the ones that we happen to catch in the act [see,
for example, \citet{elb95}].

However, there is a problem with this scenario.  \citet{bsdmm89}
showed that elliptical galaxies that are active, as determined by
their powerful radio emission, are structurally different from
inactive elliptical galaxies.  In particular, they showed that
radio-loud ellipticals typically have boxy isophotes, while
radio-quiet systems have more disk-like distortions in their
isophotes.  Such boxy isophotes are created in major mergers
\citep{nbh99}, which suggests that this violent process might form a
suitable triggering mechanism for AGN activity.  However, \citet{kh00}
showed that such a mechanism would require the subsequent lifetime of
the AGN to be only $\sim 10^7\,{\rm years}$ to explain the observed
frequency of AGN.  The secular evolution timescale for an elliptical
galaxy to lose its boxy isophotes is very much longer, so one would
therefore expect to find a large population of radio-quiet ellipticals
that have passed their active phase yet still have boxy isophotes.  The
absence of such a population in the study by \citet{bsdmm89} implies
that there must be more to nuclear activity than a simple, universal
triggering mechanism.

Indeed, one can come up with a plausible argument as to why nuclear
activity might be more generally connected to the large-scale isophote
shapes of the host galaxy.  Disky isophotes in ellipticals are
generally associated with significant rotation, and a galaxy with such
strong rotation will contain more material following nearly
circular orbits, which avoid the centre of the galaxy.  Thus, such
systems will not have as ready a fuel supply for the nucleus,
explaining the preponderance of active nuclei in systems with boxy
isophotes.

Although this connection between galaxy structure and nuclear activity
complicates the story somewhat, it is thus far fairly limited in its
impact.  The powerful radio sources identified in \citet{bsdmm89} are
almost all hosted by very luminous ($M_B < -21.8$) elliptical
galaxies.  Since such systems are often the brightest cluster galaxies
that lie right at the centres of clusters, they are rather exceptional
objects.  It should therefore not be particularly surprising if their
evolution and the origins of their nuclear activity differed from
those of more run-of-the-mill ellipticals.  It is hence still
possible that generic nuclear activity is not related to the
larger-scale properties of the host galaxy, but arises from a simpler,
essentially random, fueling process.

In this paper, we set out to see whether this stochastic scenario
remains credible, by looking for similar correlations between nuclear
activity and galaxy structure in fainter elliptical systems with
lower levels of AGN activity.  Section~\ref{sec:data} describes the
archival data upon which this analysis is based, and
Section~\ref{sec:results} presents the results.
Section~\ref{sec:disc} briefly discusses their implications.

\section{The Data}\label{sec:data}

The most thorough search for weak nuclear activity was undertaken by
\citet{hfs95}, who obtained spectra of the nuclei of a
magnitude-limited sample of almost 500 galaxies, of which 57 were
classified as ellipticals.  The emission line strengths of these
galaxies were analyzed in \citet{hfs97}, where they were classified as
inactive or active, depending on whether nuclear emission lines were
detected.  The active nuclei were further subdivided into H{\sc ii}
nuclei, transition objects, LINERs and Seyfert nuclei, with these
classifications in increasing order of excitation.  Of the 57
ellipticals under consideration here, 27 were classified as inactive,
5 as transition objects, 21 as LINERs, and 4 as Seyferts.  For this
study we adopt these classifications as indicators of nuclear
activity.  We also use the values for B-band absolute
magnitude, $M_B$, given in \citet{hfs97}.

If the argument that AGN activity is suppressed by rotational motion
is valid for these galaxies, one might expect to find a correlation
between the nuclear classifications and the larger-scale dynamical
parameters of these galaxies.  No homogeneous study has been made of
the kinematics of this sample, but the existing long-slit spectral
data on a large number of galaxies has been very helpfully collected
in the HyperLeda database \citep{ps04}, which provides values for the
central velocity dispersion, $\sigma$, and maximum rotation speed,
$v$, as derived from all archived long-slit observations along each
galaxy's major axis.

Unfortunately, the data in the HyperLeda archive is of somewhat
variable quality, so we have adopted the following protocol to extract
`best'' estimates for the requisite kinematic parameters.  Where
there is only one major-axis observation in the database, we use it.
Where there are two observations, we use the one with the smaller
quoted error.  Where there are more than two observations, we take the
value with the smallest quoted error: if this number is consistent with
the mean and standard error of the remaining observations, we adopt it
as the best estimate; if it is inconsistent with the other
observations, we reject it and repeat the process.  This approach
tends to pick out the latest, highest-quality data, rather than
averaging good data with bad, but also weeds out the relatively small
number of cases where the latest measurement is inconsistent with a
consensus of previous observations.  Although it is unlikely that this
procedure excludes absolutely all bad data, it will not preferentially
affect either the active or the inactive systems, so should not
prejudice this statistical comparison between the two types.  Of the 57
elliptical galaxies in the \citet{hfs97} sample, the requisite
kinematic data are available for all except four, reducing the
final sample size to 53 objects.

To assess the significance of the amount of rotation in an elliptical
galaxy, we also need to know its observed ellipticity, $\epsilon$.  We
obtained these values from the 2MASS extended source database, using
the published axis ratio measured from the J+H+K super image
(``sup\_ba''), as accessed through the NASA Extragalactic Database
(NED).\footnote{The NASA/IPAC Extragalactic Database (NED) is operated
by the Jet Propulsion Laboratory, California Institute of Technology,
under contract with the National Aeronautics and Space
Administration.}  

\begin{table}
\centering
\begin{minipage}{140mm}
  \caption{Final data set.\label{tab:data}}
  \begin{tabular}{@{}lcrrcc@{}}
  \hline
Name   & AGN&$v$\phantom{xx}&$\sigma\phantom{xx}$&$\epsilon$&$M_B$\\ 
       & \phantom{x}Class\footnote{0 = inactive, 1 = transition object, 2 = LINER, 3 = Seyfert}
            &(km/s)&(km/s)\\  
\hline
NGC~147 & 0 & 6 & 23 & 0.693 & $-14.52$ \\
NGC~185 & 3 & 3 & 25 & 0.891 & $-14.95$ \\
NGC~205 & 0 & 3 & 30 & 0.594 & $-15.44$ \\
NGC~221 & 0 & 45 & 80 & 0.913 & $-15.51$ \\
NGC~315 & 2 & 32 & 336 & 0.760 & $-22.22$ \\
NGC~410 & 1 & 40 & 300 & 0.740 & $-22.01$ \\
NGC~777 & 3 & 35 & 295 & 0.900 & $-21.94$ \\
NGC~821 & 0 & 84 & 208 & 0.800 & $-20.11$ \\
NGC~1052 & 2 & 120 & 240 & 0.760 & $-19.90$ \\
NGC~2634 & 0 & 65 & 190 & 0.920 & $-19.96$ \\
NGC~2768 & 2 & 110 & 185 & 0.462 & $-21.17$ \\
NGC~2832 & 2 & 8 & 320 & 0.820 & $-22.24$ \\
NGC~3193 & 2 & 80 & 200 & 0.840 & $-20.10$ \\
NGC~3226 & 2 & 40 & 201 & 0.800 & $-19.40$ \\
NGC~3377 & 0 & 87 & 140 & 0.620 & $-18.47$ \\
NGC~3379 & 2 & 40 & 240 & 0.950 & $-19.36$ \\
NGC~3608 & 2 & 26 & 194 & 0.800 & $-20.16$ \\
NGC~3610 & 0 & 150 & 180 & 0.800 & $-20.79$ \\
NGC~3613 & 0 & 141 & 220 & 0.520 & $-20.93$ \\
NGC~3640 & 0 & 120 & 183 & 0.820 & $-20.73$ \\
NGC~4125 & 1 & 76 & 240 & 0.671 & $-21.25$ \\
NGC~4168 & 3 & 11 & 190 & 0.840 & $-19.07$ \\
NGC~4261 & 2 & 30 & 330 & 0.840 & $-21.37$ \\
NGC~4278 & 2 & 60 & 270 & 0.900 & $-18.96$ \\
NGC~4291 & 0 & 68 & 300 & 0.780 & $-20.09$ \\
NGC~4339 & 0 & 28 & 111 & 0.960 & $-19.72$ \\
NGC~4365 & 0 & 70 & 270 & 0.836 & $-20.64$ \\
NGC~4374 & 2 & 10 & 300 & 0.970 & $-21.12$ \\
NGC~4406 & 0 & 20 & 255 & 0.825 & $-21.39$ \\
NGC~4472 & 3 & 45 & 300 & 0.913 & $-21.80$ \\
NGC~4473 & 0 & 54 & 200 & 0.580 & $-20.10$ \\
NGC~4478 & 0 & 62 & 150 & 0.840 & $-18.92$ \\
NGC~4486 & 2 & 2 & 350 & 0.990 & $-21.64$ \\
NGC~4494 & 2 & 54 & 150 & 0.850 & $-19.38$ \\
NGC~4552 & 1 & 10 & 260 & 0.940 & $-20.56$ \\
NGC~4564 & 0 & 140 & 170 & 0.480 & $-19.17$ \\
NGC~4589 & 2 & 35 & 220 & 0.810 & $-20.71$ \\
NGC~4621 & 0 & 100 & 240 & 0.671 & $-20.60$ \\
NGC~4636 & 2 & 27 & 235 & 0.836 & $-20.72$ \\
NGC~4649 & 0 & 43 & 370 & 0.891 & $-21.43$ \\
NGC~4660 & 0 & 140 & 210 & 0.600 & $-19.06$ \\
NGC~5077 & 2 & 25 & 254 & 0.780 & $-20.83$ \\
NGC~5322 & 2 & 80 & 250 & 0.660 & $-21.46$ \\
NGC~5557 & 0 & 10 & 250 & 0.860 & $-21.17$ \\
NGC~5576 & 0 & 10 & 200 & 0.680 & $-20.34$ \\
NGC~5638 & 0 & 62 & 154 & 0.920 & $-20.21$ \\
NGC~5813 & 2 & 90 & 235 & 0.730 & $-20.85$ \\
NGC~5831 & 0 & 27 & 152 & 0.920 & $-19.96$ \\
NGC~5846 & 1 & 60 & 250 & 0.920 & $-21.36$ \\
NGC~5982 & 2 & 80 & 255 & 0.700 & $-20.89$ \\
NGC~6702 & 2 & 20 & 172 & 0.760 & $-20.95$ \\
NGC~7619 & 0 & 63 & 325 & 0.820 & $-21.60$ \\
NGC~7626 & 2 & 25 & 260 & 0.880 & $-21.23$ \\
\hline
\end{tabular}
\end{minipage}
\end{table}

The final combined data set of galaxy names, nuclear activity
classifications, maximum rotation velocities, central velocity
dispersions, ellipticities and absolute magnitudes is presented in
Table~\ref{tab:data}.

\section{Results}\label{sec:results}

The simplest diagnostic of the dynamics of elliptical galaxies
involves a plot of $v/\sigma$ against $\epsilon$.  The former quantity
provides a measure of the division between streaming and random
motions, and on the basis of a simple dynamical argument one would
expect galaxies in which this ratio is higher to be more flattened
\citep[, p.~216]{bt87}.  In fact, most galaxies rotate more slowly than
one would predict from their flattening on the basis of the simplest
dynamical model \citep{defis83}, which is taken as evidence that most
ellipticals are not flattened by rotation.  Nonetheless, such a plot
still forms a very useful diagnostic.  

\begin{figure}
\includegraphics[width=84mm]{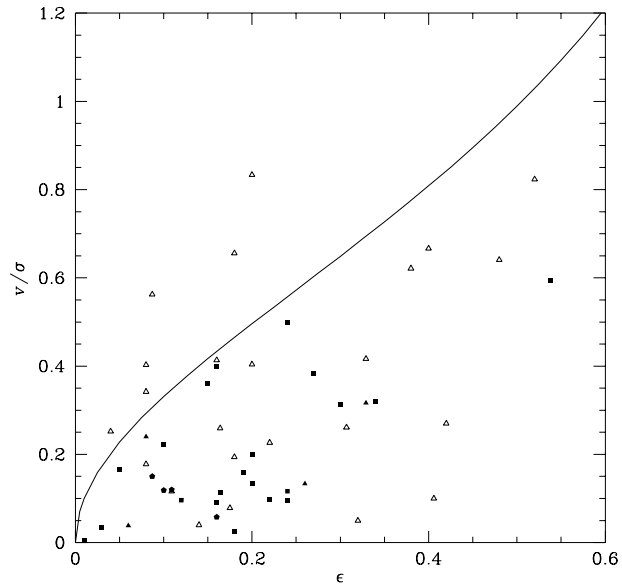}
\caption{
Plot showing degree of rotational support as quantified by $v/\sigma$
as a function of the observed ellipticity of the galaxy.  Open
triangles are inactive galaxies, while filled points represent
galaxies with various classes of active nucleus: filled triangles are
transition objects, filled squares are LINERs and filled pentagons are 
Seyferts, as classified by \citet{hfs97}.  The line shows the
predicted curve for an oblate isotropic rotator.
}
\label{fig:voversig}
\end{figure}

Figure~\ref{fig:voversig} shows the data described in
Section~\ref{sec:data} presented in this way.  As previously
mentioned, the vast majority of galaxies rotate more slowly than the
prediction of the simple isotropic rotator model.  However, it is
notable that a few galaxies are ``super-rotators,'' lying above the
curve in Fig.~\ref{fig:voversig}.  At first glance, these galaxies
seem problematic, since a rather contrived pure elliptical galaxy
dynamical model would be required to populate this part of the plane.
However, there is a simple explanation: the observed ellipticity is a
truly global measure of the properties of the galaxy, whereas the
nature of the spectroscopy means that the kinematic measurements only
probe the galaxy close to its major axis.  Thus, a relatively modest
disk component embedded in an elliptical would have little effect on
the measured ellipticity, but could dominate the measured kinematics,
making the galaxy appear to rotate faster than its global ellipticity
would allow.  Alternatively, it is possible that some of these systems
are merger remnants that are still evolving, in which case, one would
not necessarily expect to find them in the parts of
Fig.~\ref{fig:voversig} populated by equilibrium models.

Although these factors complicate the dynamical interpretation of any
observed value of $v/\sigma$, one can still use it to make a
comparison between populations: if active and inactive nuclei lie in
host galaxies with indistinguishable dynamical properties, then the
observed distribution of values of $v/\sigma$ should be identical for
both populations.  It is therefore notable that the two distributions
do not seem to be the same.  In particular, the super-rotators in
Fig.~\ref{fig:voversig} are all inactive galaxies.  This difference
suggests that the moderate AGN in this sample follow a similar trend
to the luminous radio galaxies: a large amount of rotation seems to
suppress nuclear activity.  However, it is not legitimate to pick out
regions of this plot {\it a posteriori} and claim a statistically
significant result; rather, one must look at the entire sample without
such prejudice, and compare galaxies with differing ellipticities in a
consistent objective manner.  To this end, one can calculate the
quantity $(v/\sigma)^*$ by dividing the observed value of $v/\sigma$
by the value one would obtain for an isotropic rotator of the same
ellipticity \citep{k82}.  Thus, galaxies with values of $(v/\sigma)^*
\sim 1$ are strongly rotationally supported, while for those with
$(v/\sigma)^* \sim 0$ rotation is not dynamically significant.

\begin{figure}
\includegraphics[width=84mm]{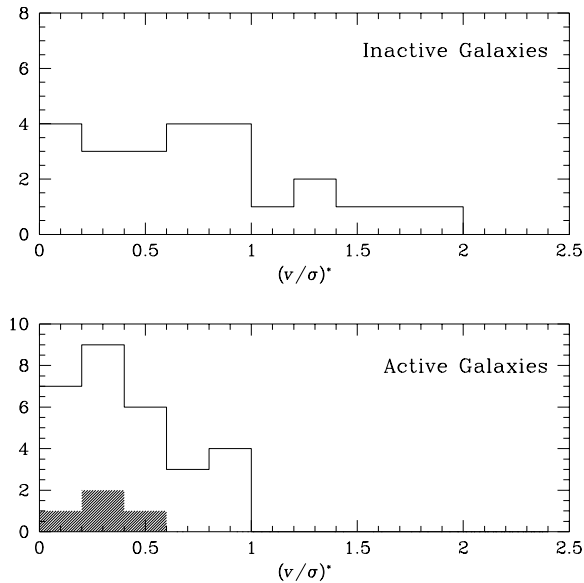}
\caption{
Histograms of the degree of rotational support in inactive and active
galaxies, as quantified by $(v/\sigma)^*$.  The shaded section of the
lower histogram indicates the highest ionization state active
galaxies, which have been classified as Seyferts.
}
\label{fig:voversigstar}
\end{figure}

The distribution of this dimensionless rotation parameter for the
active and inactive galaxies is shown in Fig.~\ref{fig:voversigstar}.
These distributions confirm the qualitative impression of
Fig.~\ref{fig:voversig}: there is an excess of rapidly-rotating
inactive galaxies.  A KS test shows that these two distributions are
different at $>95\%$ confidence.  There is even some indication of a
graduation in the degree of activity with the degree of rotational
support: the highest ionization-state ``Seyfert'' nuclei, highlighted
in the lower panel of Fig.~\ref{fig:voversigstar} seem to lie
systematically at even lower values of $(v/\sigma)^*$.  However, the
size of the sample is too small to establish such a result at a
statistically significant level.

One remaining question is whether the active and inactive galaxies
differ in some other systematic way that might be the fundamental
driver of both their activity and degree of rotational support.  In
particular, it is well established that brighter ellipticals tend
to be less rotationally supported \citep{defis83}.  If these brighter
galaxies also have systematically higher levels of nuclear activity,
then luminosity might be the underlying cause of both phenomena.

\begin{figure}
\includegraphics[width=84mm]{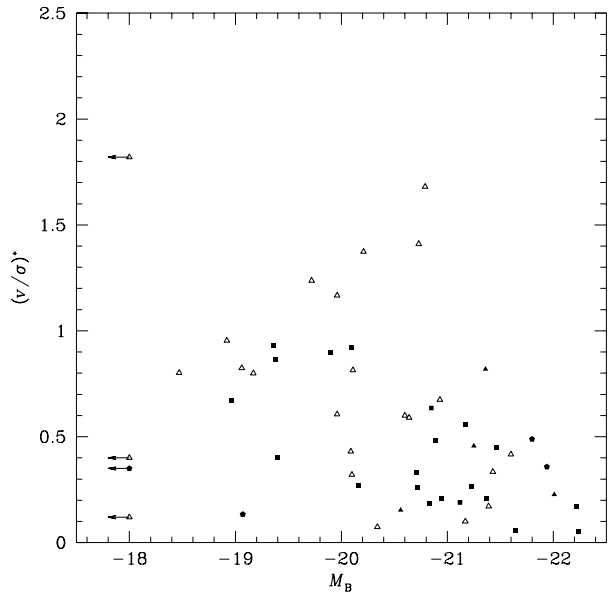}
\caption{
Degree of rotational support as a function of the absolute magnitude
of the galaxy, with open and filled symbols representing inactive and
active galaxies as in Fig.~\ref{fig:voversig}.  The dwarf galaxies are
plotted at $M_B = -18$ to keep this axis on a sensible scale.
}
\label{fig:voversigmag}
\end{figure}

Figure~\ref{fig:voversigmag} shows a plot of $(v/\sigma)^*$ against
the absolute magnitudes of the galaxies.  This plot confirms the
absence of rotational support in the brightest galaxies with $M_B <
-21$.  It further suggests that the faintest galaxies also tend to
lack rotational support, as first pointed out by \citet{bn90},
although for these lower luminosity galaxies it is also likely that
the kinematic data do not reach to large enough radii to measure the
true maximum rotation speed \citep{ps03}.  However, even if there are
such systematic biases in the observed values of $(v/\sigma)^*$, they
should affect active and inactive galaxies in the same way, so they
should have no impact on the comparisons made here.

There is also a clear indication in Fig.~\ref{fig:voversigmag} that
the most luminous galaxies are more likely to be active, with all of
the systems with $M_B < -21.6$ classified as AGN.  However, even if
one takes the extreme measure of excluding these six galaxies from the
sample, the distributions of $(v/\sigma)^*$ for the active and
inactive galaxies remain different at $> 95\%$ confidence in a KS
test.  Thus it would appear that, even if one excludes the AGN hosted
by the most massive galaxies, there are systematic differences between
the dynamics of active and inactive galaxies.

It would be desirable to subdivide the data further in absolute
magnitude to look for subtler trends with luminosity.  For example,
the referee considered the subsample at $-20.5 > M_B > -21.5$ and
pointed out that, if one excludes the two outlying rapid-rotators,
there is little to distinguish between active and inactive galaxies.
Unfortunately, the data set is really too small to draw any strong
statistical conclusions from such subsamples.  It is, however, notable
that the upper envelope of the fastest rotating galaxies fainter than
$M_B = -21$ seems to consist almost entirely of inactive galaxies.  At
the very least one can say that, independent of luminosity, there is
strong evidence that the most rapidly rotating galaxies are
systematically less active than the slower rotators.

\section{Discussion}\label{sec:disc}

The relationship between nuclear activity and the dynamics of
elliptical galaxies, originally established for luminous radio sources
in the brightest systems, seems to extend to lower levels of AGN
activity in fainter, more ordinary ellipticals.  This connection
suggests that there is more to AGN activity in ellipticals than a
universal stochastic triggering mechanism.

One possibility that we have begun to explore with this data set is
that the underlying common factor could be the luminosity (or mass) of
the galaxy.  Indeed, the most luminous galaxies are all active and
tend to rotate slowly.  Thus, it is possible that luminosity might be
a controlling parameter: more luminous galaxies contain more massive
black holes that are more likely to be active, and perhaps the
formation mechanism for these very bright systems tends to suppress
their rotation.  However, even if this is the case, it is not
inconsistent with large-scale dynamics playing a critical role in
dictating the nuclear activity of a galaxy.  If the formation
mechanism for elliptical galaxies tends to produce a lower value of
$(v/\sigma)^*$ for a more luminous galaxy, then such a luminous system
will contain more material on orbits that travel close to the centre
of the galaxy.  This ready fuel supply will enhance the mass of the
central black hole and its current accretion rate, explaining why such
a luminous galaxy is so likely to contain an active nucleus.

Quite which mechanisms are the fundamental drivers of AGN activity
remains to be established beyond all doubt.  However, hopefully this
paper has demonstrated that the large-scale dynamics of the galaxy is
likely to be a significant factor.  In retrospect, this really should
not be a surprise: galaxies are fundamentally dynamical entities, and
any model of AGN fueling that glosses over this point cannot provide
an accurate description; just because two elliptical galaxies look
similar when they are photographed, it does not mean that their inner
workings are the same, so there is no reason why both should have the
same probability of containing an active nucleus.

\section*{Acknowledgments}
I am most grateful to the referee, Philippe Prugniel, for his insights
into the importance of luminosity as a parameter when studying
nuclear activity and large-scale dynamics.

\label{lastpage}

\end{document}